# Energy Landscape of Sugar Conformation Controls the Sol-to-Gel Transition in Self-assembled Bola Glycolipid Hydrogels

Alexandre Poirier,[a] Patrick Le Griel,[a] Thomas Zinn,[b] Petra Pernot,[b] Sophie L. K. W. Roelants,[c,d] Wim Soetaert,[c,d] Niki Baccile[a*]

[a] Sorbonne Université, Centre National de la Recherche Scientifique, Laboratoire de Chimie de la Matière Condensée de Paris, LCMCP, F-75005 Paris, France

[b] ESRF – The European Synchrotron, CS40220, 38043 Grenoble, France

[c] Centre for Industrial Biotechnology and Biocatalysis (InBio.be), Department of Biotechnology, Faculty of Bioscience Engineering, Ghent University, Coupure Links 653, 9000 Ghent, Belgium

[d] Bio Base Europe Pilot Plant, Rodenhuizenkaai 1, 9042 Ghent, Belgium

* Corresponding author:
Dr. Niki Baccile
E-mail address: niki.baccile@sorbonne-universite.fr
Phone: +33 1 44 27 56 77

**Abstract**

Self-assembled fibrillar network (SAFIN) hydrogels and organogels are commonly obtained by a crystallization process into fibers induced by external stimuli like temperature or pH. The gel-to-sol-to-gel transition is generally readily reversible and the change rate of the stimulus determines the fiber homogeneity and eventual elastic properties of the gels. However, recent work shows that in some specific cases, fibrillation occurs for a given molecular conformation and the sol-to-gel transition depends on the relative energetic stability of one conformation over the other, and not on the rate of change of the stimuli. We observe such a phenomenon on a class of bolaform glycolipids, sophorosides, similar to the well-known sophorolipid biosurfactants, but composed of two symmetric sophorose units. A combination of oscillatory rheology, small-angle X-ray scattering (SAXS) cryogenic transmission electron microscopy (cryo-TEM) and *in situ* rheo-SAXS using synchrotron radiation shows that below 14°C, twisted nanofibers are the thermodynamic phase. Between 14°C and about 33°C, nanofibers coexist with micelles and a strong hydrogel forms, the sol-to-gel transition being readily reversible in this temperature range. However, above the annealing temperature of about 40°C, the micelle morphology becomes kinetically-trapped over hours, even upon cooling, whichever the rate, to 4°C. A combination of solution and solid-state nuclear magnetic resonance (NMR) suggests two different conformations of the 1'', 1' and 2' carbon stereocenters of sophorose, precisely at the β(1,2) glycosidic bond, for which several combinations of the dihedral angles are known to provide at least three energetic minima of comparable magnitude and each corresponding to a given sophorose conformation.

**Introduction**

Fibrillation and hydrogelation from low molecular weight (LMW) compounds is a deeply studied topic in the field of soft matter, for the interesting and high-end applicative perspectives in fields as wide as tissues engineering, pharmacology or art preservation.[1–4] The large body of work performed in the past three decades has largely focused on the discovery of new LMW hydrogelators and control of the fibrillation and hydrogel properties.[5–12] Only recently, however, the interest began to focus on the relationship between the energy landscapes of a given gelator and its self-assembled structures. It has been shown for peptide amphiphiles that a relatively classic scheme of intermolecular interactions, temperature annealing and ionic strength can stabilize specific molecular conformations in a kinetically trapped state, with the consequence of stabilizing fibrillar structures, otherwise poorly stable.[13] A more complex energy landscape panorama has also been considered to explain the larger number of coexisting supramolecular morphologies (spheres, ribbons, tubes) for Fmoc-derivatives.[14] Other groups have worked on the relationship between molecular conformation and gel properties, even without mentioning energy landscapes in clear.[11,12] Despite these and few other examples, this approach, consisting in connecting specific, energetically-stable, molecular conformations with supramolecular self-assembly and, eventually, macroscale properties, like hydrogelling, are relatively rare.

This relationship is unexpectedly shown here for a new class of amphiphiles of biological origin. Widely known as biosurfactants, this family of compounds, produced by a fermentative approach using microorganisms in the presence of sugars and fatty acids,[15,16] was initially developed for the field of "green" detergency, with more recent applications in environmental science, oil recovery and cosmetics.[16–20] However, more recent research also shows that the word biosurfactant is highly reductive, as most of these molecules, under well-controlled physicochemical conditions, can form interesting self-assembled structures and soft materials with unexpected complexity,[21] such as a rare class of hydrogels with lamellar structure,[22] or more classical fibrillar hydrogels, but with unique elastic properties depending on the pH-change rate.[23]

Stringent regulations combined to more and more demanding consumers strongly drive the research interest of private companies[24] towards the development of safe-by-design products. Within this context, the development and understanding of soft materials from biological amphiphiles constitutes a new research trend with high priority, due to the application potential of safe soft materials.

Despite the similar fibrillation and gelation properties between some microbial amphiphiles and well-established LMW compounds, like FMOC derivatives,[9,25] recent work has shown that microbial amphiphiles can have a complex, hard to predict, behaviour. For instance, C18:0 derivatives of sophorolipid biosurfactants fibrillate and form hydrogels under slow pH change rates,[23] while gelation of the C16:0 derivative of the same compound is independent of the pH variation rate, but rather affected by the final pH of the gel.[26] Interestingly, the latter is more coherent with the literature on pH-dependent low-molecular weight gelators,[27,28] while the former followed a path, generally observed for temperature-driven LMW hydro- and organogelators.[29]

In a previous work, we studied the self-assembly of a symmetrical bolaamphiphile sophoroside ($SS_{bola}C18:1$, Figure 1) in water employing synchrotron SAXS analysis on freshly prepared samples. $SS_{bola}C18:1$ was shown to form a stable micellar phase, size and shape of which are almost invariant with concentration up to about 10 wt%.[30] More recent experiments have however shown the property of freshly prepared aqueous $SS_{bola}C18:1$ samples to form fibrous hydrogels, contradicting previous data, at first glance. In this work, we invoke an energy landscape approach to explain the strong fibrillation and hydrogelation kinetics of a symmetrical bolaamphiphile sophoroside (Figure 1). Previously reported to form a micellar phase up to at least 10 wt% in water,[30] this compound actually undergoes a slow micelle-to-fiber, sol-to-gel, transition at room temperature. Micelles (sol) and fibers (gel) are the thermodynamically stable phase above 40°C and below 14°C, respectively. Interestingly, the micellar sol becomes kinetically trapped during hours and days, even upon cooling below 40°C and at temperatures as low as 4°C. The sol-to-gel transition is instead readily reversible, within the order of minutes, when the gel is heated and cooled just below 40°C.

Recently reported for amphiphilic peptides, the kinetically trapped and thermodynamic states were respectively associated to transient and stable conformations of the peptidic moiety.[13] Here, a combination of solution and solid-state $^{13}C$ nuclear magnetic resonance (NMR) show that two different conformations of the sophorose headgroup are associated to the fibrillar and micellar structures, respectively. The spectroscopic changes (chemical shifts) only involve the signal of the carbon stereocenters on the β(1,2) linkage of sophorose and this is explained by the well-known variations in the dihedral angles, φ and ψ, around the glycosidic linkage in disaccharides, generating minima of conformational energy, with unique NMR (J-coupling, chemical shift) parameters, both in solution and solid-state.[31–35] Two energetic minima of comparable magnitude, each associated to a specific conformation of sophorose with a distinct pair of φ and ψ, can then be associated to the micellar and fibrillar structures. In this regard, the relationship between sugar conformation, structure and NMR has been reported for solid hexosamides amphiphiles.[36–38]

The results shown here for the symmetrical bolaamphiphile sophoroside could actually help understanding the phase behaviour of other biobased glycolipid amphiphiles, for which sugar conformation effects on their properties are still unknown. Looking at conformational issues of the sugar could help predicting the properties, so to better define a possible application of biosurfactant-based soft materials.

**Material and methods**

*Chemicals, sample preparation.* The monounsaturated symmetrical sophoroside SS$_{bola}$C18:1 ($M_w$ = 933.0 g.mol$^{-1}$) contains a nonacetylated sophorose group at each extremity of the aliphatic chain. The molecule is obtained by fermentation from of the genetically-modified yeast *Starmerella bombicola*. The protocol is described in Ref. [30] and SS$_{bola}$C18:1 corresponds to the molecule presented in Figure 1e of Ref [30] (oleyl alcohol based symmetrical bolaform (sBola) sophorosides (SSs) (C18:1)). The compound is provided by Bio Base Europe Pilot Plant, Gent, Belgium, under the name non acetylated symmetrical bola sophoroside (C18:1, w), lot N° CBS_SL42 Inv63. The compound is composed of 98.7 % of SS$_{bola}$C18:1 (HPLC-ELSD analysis), while $^1$H and $^{13}$C NMR analyses are given in Figure S23 (compound XXX1) and Tables SVII, SVIII of Ref. [30]. The product is used without further purification. The biosurfactant is mixed in milliQ water at room temperature (RT).

*Rheology.* A MCR 302 rheometer (Anton Paar, Graz, Austria) is used with cone-plate geometry (Ø: 50mm, gap at center 0.210 mm) at a regulated temperature. Solvent trap with water is used to minimize evaporation. ~ 1.3 mL of gel is loaded on the center of the plate using a spatula to prevent bubbles, then the excess is removed. Value of the pseudo-equilibrium G' is taken after 5 min of oscillatory measurement at 1 Hz and low strain γ, one order of magnitude lower than the critical strain.

*Cryogenic-transmission electron microscopy (Cryo-TEM).* Pictures are recorded on an FEI Tecnai 120 twin microscope operating at 120 kV with an Orius 1000 CCD numeric camera. The sample holder is a Gatan Cryo holder (Gatan 626DH, Gatan). Digital Micrograph software is used for image acquisition. Cryo-fixation is done with low dose on a homemade cryo-fixation device. The solutions are deposited on a glow-discharged holey carbon coated TEM nikel grid (Quantifoil R2/2, Germany). Excess solution is removed and the grid is immediately plunged into liquid ethane at -180°C before transferring them into liquid nitrogen. All grids are kept at liquid nitrogen temperature throughout all experimentation. Cryo-TEM images are treated and analyzed using Fiji software, available free of charge at the developer's web site.[39]

*Small Angle X-ray Scattering (SAXS).*

*SAXS.* SAXS experiments on selected samples (Figure 1) were performed on the BioSAXS BM29 beamline at the European synchrotron, ESRF-EBS[40] (Experiment: MX-2311, Grenoble, France) using X-ray radiation with energy of 12.5 keV and the sample-detector distance of 2.867 m, corresponding to the beamline standard configuration. The photon energy is calibrated by measuring the $L_I$ and $L_{III}$ edges of platinum and the sample-to-detector distance is determined using silver behenate ($d_{ref}$ = 58.38 Å).[42,43] Viscous samples and gels are loaded manually using a 1 mL syringe, into a 1 mm quartz glass capillary. Samples are manually loaded in the capillary using a syringe. The signal of the Pilatus 2M 2D detector, used to record the data, was integrated azimuthally with PyFAI to obtain the scattered intensity *I(q) vs. q* spectrum ($q = 4\pi \sin\theta / \lambda$, where 2θ is the scattering angle) after masking systematically wrong pixels and the beam stop shadow.[415]

*Rheo-SAXS.* Experiments coupling rheology to SAXS is performed on the ID02 beamline at the ESRF synchrotron (Grenoble, France) during the proposal N° SC-4976. The energy of the beam is set at 12.28 KeV and the sample-to-detector distance at 1.5 m. The beamline is equipped with a Haake Rheo-Stress RS6000 stress-controlled rheometer containing a polycarbonate couette cell having a gap of 0.5 mm and a required sample volume of 3 mL. The rheometer is controlled through an external computer in the experimental hutch using the software Rheo-Win. The temperature of the cell is set at 25°C, unless otherwise stated. Experiments are performed in a radial configuration. The absolute intensity is obtained by subtracting the background (polycarbonate cell containing milliQ water) and by dividing the signal by the effective thickness of the sample in cell (0.17 cm). The SAXS acquisitions are manually triggered at the same time as the rheology acquisition, with an error of ± 2 s. The frequency of data recording is manually set and determined independently for both SAXS and rheology measurement. Shear and oscillatory measurement are performed at various velocity, strain and frequency.

*Nuclear magnetic resonance (NMR).* $^1$H solution-state NMR measurements were performed on a Bruker Avance II spectrometer ($^1$H Larmor frequency of 400.2 MHz) equipped with a BBFO 5 mm probe. Single-pulse experiments were carried out using a 90°($^1$H) pulse length of 10.75 µs and a recycle delay of 3 s. All the samples were analysed in neat $D_2O$.

$^{13}$C solid-state NMR experiments were carried out on an Avance III HD Bruker 7.05 T ($\nu_{1H}$= 300 MHz) spectrometer and a 4 mm magic angle spinning (MAS) probe. The high-resolution $^{13}$C NMR spectra were obtained by $^1$H → $^{13}$C cross-polarization (CP) under magic angle spinning (MAS). $\nu_{MAS}$= 8 kHz, number of transients, NS= 3072; time domain size, TD= 1 k; pulse length, p($^1$H)= 2.47 µs; relaxation delay, D= 5 s; contact time during CP, $t_c$= 1000 µs; $^1$H RF field strength for high-power continuous wave decoupling at 101 kHz.

**Results**

*$SS_{bola}C18:1$ forms SAFiN hydrogels in water*

The gel properties of $SS_{bola}C18:1$ in water are put in evidence by strain sweep and frequency sweep experiments. The linear viscoelastic regime (LVER) is measured by strain sweep experiments Figure S 1a, while the appropriate strain (0.1%) is chosen about one order of magnitude under the critical strain. The frequency sweep experiment (Figure S 1b) confirms the solid-like behaviour of the network by the weak frequency dependence on the elastic and viscous moduli and the elastic modulus G´ being about one order of magnitude higher than the viscous modulus G´´.

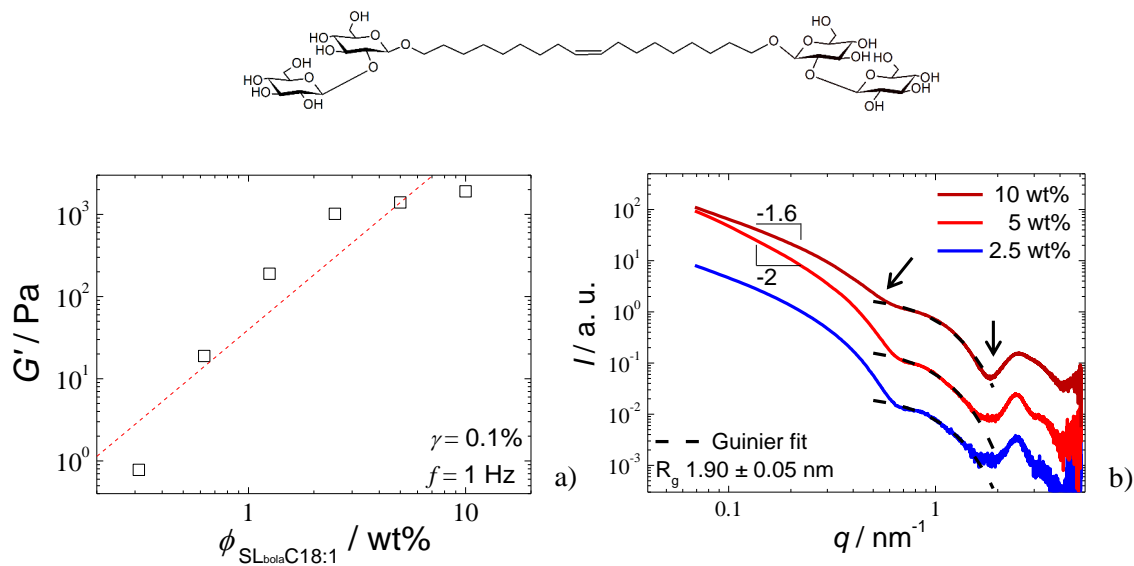

**Figure 1 -** a) Elastic moduli as a function of weight fraction, φ, of SS$_{bola}$C18:1. Straight dash line corresponds to the best fit (log-log scale). b) SAXS spectra of SS$_{bola}$C18:1 hydrogel at 2.5 wt%, 5 wt% and 10 wt%. The arrows indicate minimum of the form factor, $q_{min}$ (q> 1 nm$^{-1}$) associated to the micellar profile and the breaking poink between the micellar and ribbon signals (q ~0.6 nm$^{-1}$). Dash line corresponds to the Guinier fit, giving a gyration radius $R_g$ = 1.90 ± 0.05 nm. Data are shifted by a factor of about 10.

The elastic properties of SS$_{bola}$C18:1 hydrogels increase with its weight fraction (Figure 1a). Straight dot line indicates a power law fitting with a scaling law of 2.2 (log-log scale), typical of fibrillar entanglement in a good solvent,[42] and found before for other biosurfactant-based hydrogels.[43] Nevertheless, the mismatch between the experimental data and the power law suggests a network structure of higher complexity, as it can be deduced from the SAXS spectra in Figure 1b. At low-q, the spectra follow an approximate slope of -1.6, typically observed for elongated micelles or fractal systems,[44,45] the middle-q region exhibits the typical oscillation profile of a micellar morphology,[45] having a gyration radius $R_g$ = 1.90 ± 0.05 nm according to the Guinier fit (Figure 1b). However, at high-q, data display an oscillation of the form factor, superimposed to a correlation peak at 2.47 nm$^{-1}$, which reflects a repeated distance of 2.54 nm. The presence of such a peak in biosurfactants[26,46] and other self-assembled LMW systems[47,48] is generally associated to a semi-crystalline order inside the fibers.[48] The fibrillar structure is confirmed by complementary cryo-TEM experiments (Figure 2), showing the presence of twisted ribbons, having a typical width of 20 nm and an inter-node of about 140 nm (additional pictures Figure S 2).

SS$_{bola}$C18:1 is composed of two sophorose groups, a disaccharide of which the size is generally estimated to about 1 nm,[49] and a C18:1 hydrophobic spacer, of which the length can be reasonably estimated between 2.0 and 2.5 nm using the Tanford formula. According to the latter, L= 0.154 +$n \cdot$0.1265,[50] $n$ being the number carbon in the alkyl chain. For $SS_{bola}C18:1$, $n$= 18 and L= 2.4 nm. However, $SS_{bola}C18:1$ is monounsaturated, and the actual length may be smaller. Altogether, $SS_{bola}C18:1$ is expected to have a length in the order of, or slightly above, 4 nm. This value is consistent with the experimental diameter (2Rg= 3.8 nm) of the micelles, thus indicating that the micellar diameter is equivalent to the full molecular length, as expected for bolaform amphiphiles[12,51] and as found before for similar compounds.[12,43,52] Compared to the experimental repeating distance of 2.54 nm found in the crystalline fibers, one can reasonably estimate that $SS_{bola}C18:1$ undergoes a molecular tilt of 61° within the fiber, not unusual in biosurfactants[43,46,53] and bolaform amphiphiles in general.[47,54]

However, the presence of flat twisted ribbons is not in complete agreement with the full SAXS profile, as one expects a power-law dependence of I(q) at low-q rather in the order of -2 and a less marked oscillation of the form factor in the mid-q range.[46,55] The possible explanation of such a discrepancy will be discussed later after further experiments.

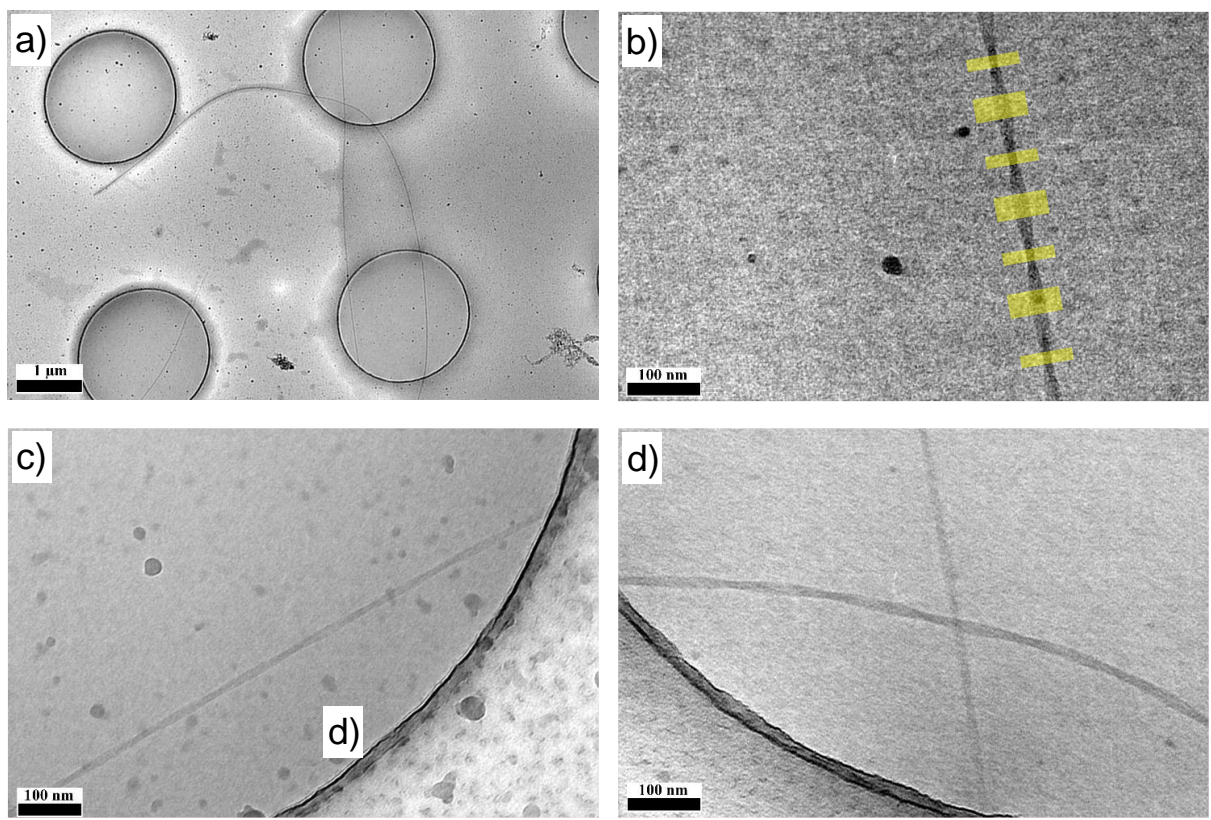

**Figure 2 - Cryo-TEM pictures of 0.05 wt% $SS_{bola}C18:1$. Yellow thick lines correspond to area used to determine fiber dimensions. Scale bars correspond to 1 μm in a) and 100 nm in b-d).**

Generally-speaking, semi-crystalline ribbons seem to be common for many neutral, acidic and basic sophorolipids SL-C18:0,[56] SL-C18:1,[53] $SS_{bola}C16:0$,[43] SL-C16:0,[26] indicating that sophorolipids may have a specific tendency to crystallize into highly anisotropic structures. However, cellobioselipids[57–59] but also glycolipids in general[54,60,61] behave in a similar way, whether in water or in organic solvents. In this regard, many other amphiphilic systems form hydrogels or organogels at low concentration and under specific temperature or pH conditions.[60–62] The presence of a sugar headgroup could be responsible for such a common behavior, and this hypothesis will be discussed in more details later on in the manuscript.

As found for other biosurfactants-based SAFiN, $SS_{bola}C18:1$ hydrogels display fast recovery after a large shear, typically few seconds after a 100% of shear strain (Figure S 3). Additionally, combined rheo-SAXS experiments in Figure S 4 show the rheo-thinning behaviour at shear rates from 0 to 100 $s^{-1}$ and they confirm the structural stability of the fibrils in the entire shear range, as shown by the superimposable SAXS profiles. In summary, $SS_{bola}C18:1$ in water forms a fibrillar gel reaching elastic moduli in the order of 1 kPa at 2 wt%. However, the SAXS curves seem to indicate a coexistence between micelles and twisted nanofibers at room temperature.

*Thermo-responsive self-assembly.*

The temperature dependence of the elastic properties is evaluated by flow behaviour in a couette cell setup coupled to the SAXS beamline giving associated structure during heating/cooling step. Rheo-SAXS experiment are shown Figure 3.

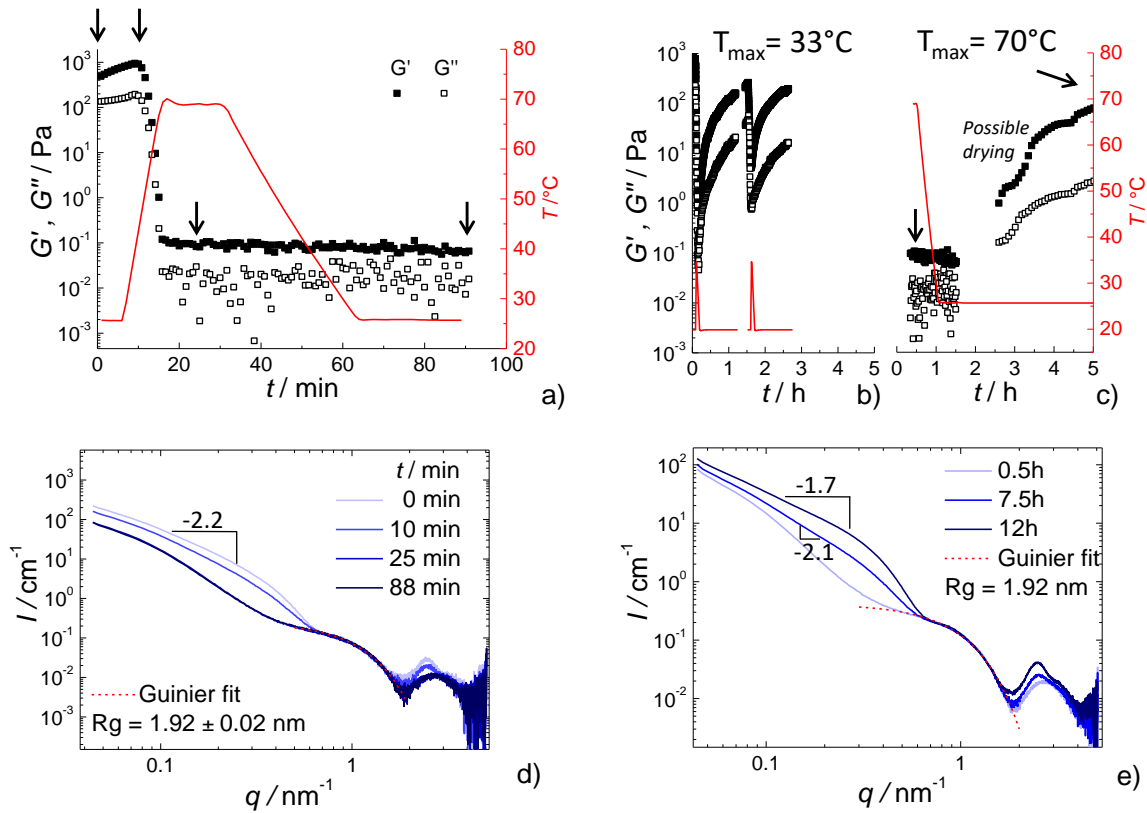

**Figure 3** – Rheo- (a) SAXS (d) experiment performed on SSbolaC18:1 (5 w%) hydrogel. Oscillatory rheology (Couette cell) is performed in the linear viscoelastic regime. Evolution of storage (G′) and loss (G′′) moduli (a) of as a function of a temperature cycle and corresponding (d) SAXS profiles (given by arrows). c, e) Rheo- (c) SAXS (e) experiment performed on SSbolaC18:1 (5 w%) hydrogel at room temperature after heating at T= 70°C. Arrows indicate the time at which SAXS profiles are extracted. Segmented red lines in d,e) corresponds to a Guinier fit with gyration radius of $R_g$= 1.92 nm. b) Rheology-only experiment (plate-plate geometry) performed on SSbolaC18:1 (5 w%) hydrogel at room temperature after heating at 33°C.

Firstly, a fast heating process is set from room temperature (RT) to 70°C, thus producing an immediate loss of the elastic properties between 35°C and 50 °C (Figure 3a). These changes reflect the gel-to-sol transition and, from a structural point view, they demonstrate a fiber-to-micelle phase transition, described before for a similar biosurfactant hydrogelator,[43] Fmoc-dipeptides[63] and other bolaform amphiphiles.[12,64] The corresponding SAXS spectra (Figure 3d) show important changes. In the low-q scattering region, the slope varies from about -2, typical for fibers, to about -3, typical of large fractal interface,[44] and probably attributable to a small fraction of large-scale objects of poorly-defined structure. The mid-q range is characteristics of a micellar signal having an unchanged Guinier radius, while the high-q region is characterized by the loss of the broad diffraction peak in favour of the first oscillation of the micellar form factor. After the gel-to-sol transition, this SAXS profile is quite invariant, both at 70°C and even after cooling back to RT (88 min, Figure 3d).

These data confirm the interpretation given above for Figure 1b, that is coexistence of fibers and micelles at RT. In the meanwhile, they also show that immediate cooling is not enough to recover the fiber morphology and to induce a sol-to-gel transition. The liquid state, associated to micelles SAXS profile is maintained over more than one hour. In fact, hydrogel formation, elasticity and the corresponding fibers are only recovered after several hours, and even days. The rheo-SAXS experiment into Figure 3c,e shows that the increasing G´ (rheology, Figure 3c) is associated to the fiber formation, shown by the low-q scattering being in the order of -2 and appearance of the broad diffraction peak at high q after at least 8-10 hours (SAXS, Figure 3e). However, one must note that the sudden increase in G' observed after about 3 h (Figure 3c) could actually correspond to a drying phenomenon in the Couette cell. This is suggested by several factors: experiments run under air during several hours are known to be affected by drying, the increase in G is too steep with respect to a classical time-dependent evolution, G' overwhelms G'' by two orders of magnitude and a number of other experiments performed in a close vial show that gels form only after few days. Interestingly, when temperature is increased only in the vicinity of 30°C, the recovery is much faster and repeatable (Figure 3b) if compared to heating at 70°C (Figure 3c), after which the recovery at room temperature occurs in the order of hours and days. This aspect will be described in more detail below.

The rheo-SAXS data are confirmed by a quantitative evaluation of the fiber-to-micelle ratio against temperature by employing solution state $^1$H NMR. This technique is sensitive only to molecules undergoing fast-tumbling motion in solution. Generally employed for dissolved compounds, it can also detect the presence of small molecular aggregates, like micelles in water. On the contrary, crystalline solids, like fibers, are too slow in the NMR time scale and their signal is not detected in the solution state. Such properties can be fruitful to measure, quantitatively, the micelle-to-fiber ratio during a sol-to-gel transition by integrating the peak area over a given spectral window.[65,66] In the present system, from rheo-SAXS experiments (Figure 3), the micellar fraction $X_M$ is found to be 100% for T = 50°C (fiber-free, micellar solution), while the fiber fraction corresponds to $X_F \equiv X_M - 1$, whereas at 4°C, $X_F$= 100% (no NMR signal).[67] The peak area can be connected to the molar fiber fraction, after normalization by the area of a pure micellar phase. Furthermore, a reduction of the molecular mobility generates broader peaks, until disappearance of the signal into the background in the limit of solids. When micelles are detected by $^1$H NMR, the width of their peak gives an idea of the micellar mobility in solution during the transition.

$^1$H solution NMR is performed on a SS$_{bola}$C18:1 sample thermalized during several days at 4°C and inserted in a pre-cooled probe. $^1$H NMR spectra are then recorded each 1°C from 4°C (hydrogel) to 50°C (solution), as shown in Figure 4a,b. Figure 4a shows that no signal is detected until 14°C, indicating a fully gelled fibrous sample, as underlined in Figure 4b. Above 14°C, a broad signal starts to appear at 1.3 ppm, corresponding to the C$\underline{H}_2$ of the aliphatic chain, indicating that the fiber-to-micelle transition takes place and that micellar diffusion is highly hindered in the gel medium. In the 14°C-39°C range, the peak intensity grows more and more while its width becomes more and more narrow, indicating generation of a larger fraction of micelles and consequent fluidification of the medium. In this range, the content of micelles settles at about 90% when temperature reaches 29°C, while an inflexion in the full-width half maximum (FWHM) is concomitantly observed, indicating a slower evolution of the molecular mobility. Above about 40°C, the micellar fraction keeps increasing between 90% and 100% while the FWHM settles at 20% of the initial width in the fiber phase.

As a first partial conclusion, the combination of rheo-SAXS and $^1$H NMR eventually shows that the loss of the elastic properties is explained by the loss of 90% of the initial fibrillary network. However, these data cannot explain a singular feature concerning the reversibility of sol-to-gel and micelle-to-fiber transitions.

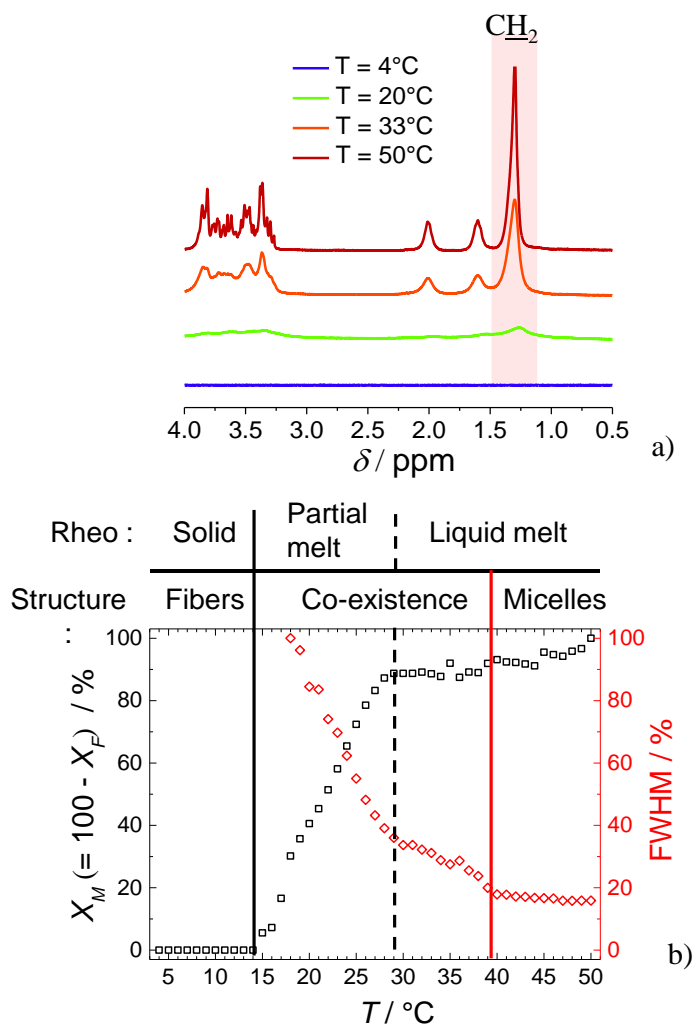

**Figure 4** – a) $^1$H solution NMR spectra of SS$_{bola}$C18:1 (C= 2.5 wt%) in D$_2$O recorded at different temperatures, each one equilibrated during 2 min. The shaded area highlights the peak assigned to the C$\underline{H}_2$ of the fatty acid at δ= 1.2 ppm and reflecting to the fraction of SS$_{bola}$C18:1 in a micellar environment. The micellar fraction, $X_M$, and the full-width half maximum (FWHM) of the C$\underline{H}_2$ peak are reported in b) as a function of temperature. $X_M$ is calculated by normalizing the C$\underline{H}_2$ peak area by its value measured at 50°C, while the fiber fraction is $X_F$= 100-$X_M$.

*Reversibility, kinetics and sugar conformation.*

In a previous work, the self-assembly of SSbolaC18:1 in water was characterized by a stable micellar morphology up to 10 wt%,[30] while this work puts in evidence a stable fibrous phase at the same temperature and in a comparable concentration range. Such puzzling result is followed by even more puzzling data recorded on the present system, where hydrogel erratically forms on fresh samples or after cooling. To better understand this phenomenon, the hypothesis is made that the final heating temperature and heating rates can play a crucial role in the reversibility of the gel-to-sol-to-gel transition. In particular, the effect of the cooling rate on the mechanical properties of SAFIN are well-known, whereas fast cooling rates promote supersaturation and spherulite formation, and the corresponding hydro- and organogels have poor elastic moduli.[5,29]

Figure 5 shows the rheological properties of three different $SS_{bola}C18:1$ hydrogels prepared at 20°C and following each a specific temperature variation profile, which allows to discriminating the effect of cooling rate (0.5 against 10°C/min) and maximum temperature, $T_{max}$. $T_{max}$ is chosen at 33°C or 50°C, the former being in the partial melt region (Figure 4) and in the proximity of the gel-to-sol transition (Figure 3a,b), and the latter being in the full micellar, sol, region. Two case scenario are shown below, when T< 40°C and T> 40°C.

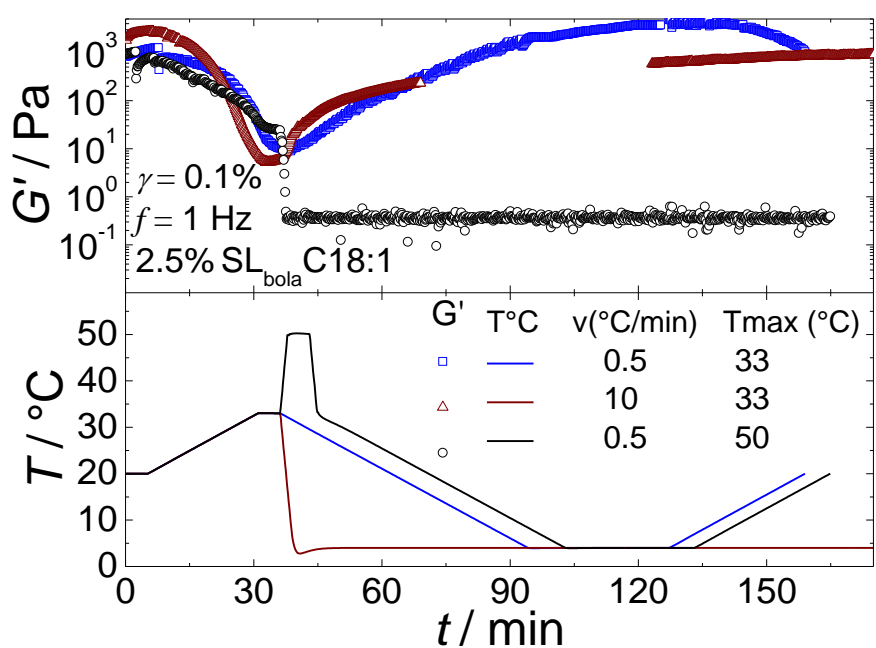

**Figure 5 – Rheological experiment. Elastic moduli of 2.5 wt% $SS_{bola}C18:1$ gel as a function of time, at different temperature rate and profile indicated on the caption.**

*T< 40°C.* When $T_{max}$ is set at 33°C, independently of the cooling rate, the hydrogels loose their elastic behaviour by at least two orders of magnitude in G´, as also described earlier (Figure 3a,b). Upon cooling to 4°C, when the fiber fraction is 100 % (Figure 4b), the elastic properties are slowly but progressively recovered within 1 to 2 hours, in agreement with previous data (Figure 3b). Nonetheless, the elastic modulus is higher by one order of magnitude when slow cooling conditions are used (green profile at 0.5°C/min against red profile at 10°C/min, Figure 5), as expected for SAFiN hydrogels. Finally, the reversibility of this process and partial loss of the elastic properties due to the gel-to-sol transition starting at 14°C (Figure 4b) is put in evidence by the loss of the elastic properties when temperature is increased again from 4°C to 20°C (green, Figure 5). The melt/growth could be repeated several time by small annealing (Figure 3b). All in all, this part of the experiment in Figure 5 shows that G´ is maximized under slow cooling rates conditions when the fiber fraction is 100%, that is below 14°C. Cooling must occur when in the partial melt region, where micelles and fibers coexist (14°C-39°C). This process is entirely reversible over several cycles (here two, but tested up to 20 for a similar bolaform sophoroside).[43]

This is commonly understood in the literature as follows. For many SAFiN, the fiber's growth is associated to the classical theory of crystallization in solution, where the kinetics of nucleation and growth is related to a modulation of the supersaturation through the solubility of the compounds.[5,29,68,69] Slow cooling rates favour nucleation over growth and reduce supersaturation, thus increasing the mismatch energy barrier. This results in fibers with a low branching degree and stronger gels. Fast cooling rates, on the contrary, favour growth over nucleation as well as supersaturation, lowering the mismatch crystallization energy and enhancing spherulite formation, hence weaker gels.[5,29]

*T> 40°C*. If, during a slow heating-cooling cycle at 0.5°C/min, similar to what is presented above, one introduces a fast heating-cooling at 50°C (black curve), the expected gel-to-sol transition is not followed by a reversible sol-to-gel process and the sample is still liquid at 4°C during several hours. This experiment demonstrates that annealing the sample at 50°C, in the full micellar region, produces a stable liquid micellar solution even after cooling and rest at 4°C during several hours, while gelation is only observed again after few days at RT (not shown). When $SS_{bola}C18:1$ is heated to the full micellar region (above about 35°C-39°C), the gel-to-sol-to-gel, that is the fiber-to-micelle-to-fiber, process is not reversible within the same time scale as when the sample is heated in the mixed micelle-fiber region.

The three temperature-dependent experiments in Figure 5 reasonably exclude the cooling kinetics as possible explanation for the partial irreversibility of the gel-to-sol-to-gel process and rather suggest the existence of a specific activation energy above which fibrillation becomes unfavourable upon cooling, may it be slow or fast. This fact excludes crystallization arguments and it rather suggests conformational differences of the $SS_{bola}C18:1$ molecule, and most likely of its sugar headgroups.

The impact of the conformation of sugar headgroups on the crystal structure and phase behavior of alkyl hexonamides has been known for a long time.[36–38] This family of compounds has a linear hexonamides headgroup, of which differences in the torsion angles were associated to given phases,[36] by associating $^{13}C$ solid-state NMR and X-ray diffraction data. Solid-state

NMR has also been used before to study the structure-relationship in bolaamphiphiles.[12] To verify the hypothesis according to which the conformation of the sophoroside headgroup is different in the micellar and fibrillar state, we perform a series of $^{13}$C CP-MAS solid-state NMR experiments on lyophilized samples, initially prepared in their full fiber (4°C), mixed fiber/micelle (20°C, 33°C) and micelle state (50°C) (Figure 6). Please note that lyophilisation of the fibers is necessary, because due to the low sample concentration (2.5 wt%) and fast spinning of the NMR rotor (8 kHz), liquid suspension of the fibers are centrifuged out onto the rotor walls, with complete loss of the signal. These data are compared to the signal of a liquid micellar solution at 55°C.

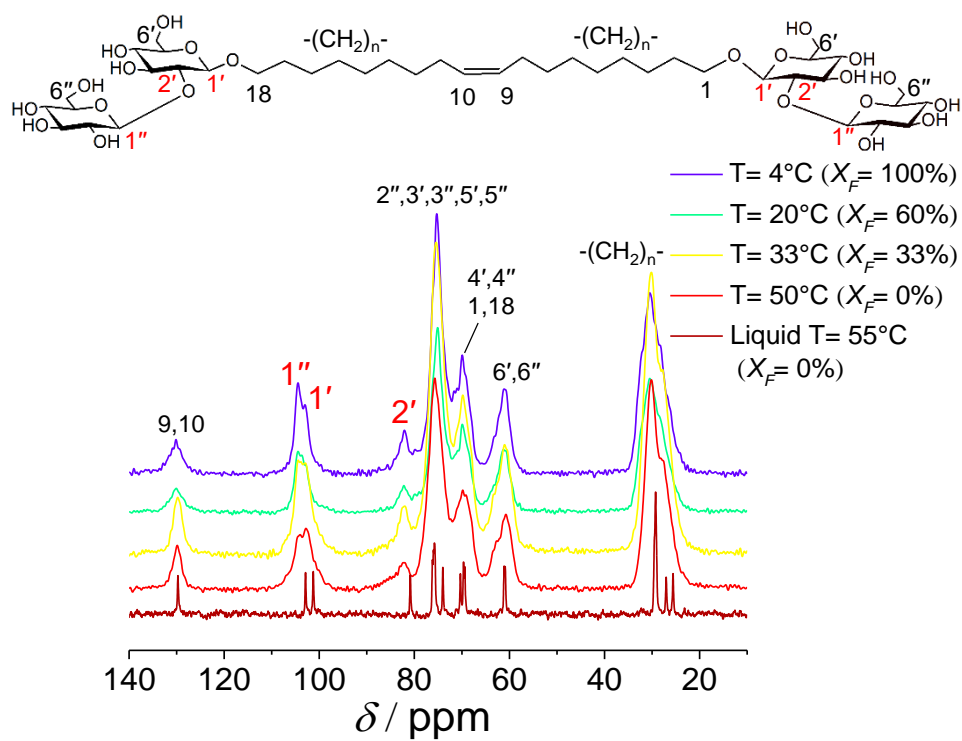

**Figure 6 – Solid-state $^{13}$C NMR spectra performed on freeze-dried solids prepared from 2.5 wt% SS$_{bola}$C18:1 solutions heat at different temperature, from 4°C to 50°C.**

Figure 6 shows the $^{13}$C solid-state CP-MAS NMR analysis of SS$_{bola}$C18:1 lyophilized samples prepared from a sol and gel state and compared to the $^{13}$C liquid NMR spectrum recorded on a micellar solution. According to the quantitative analysis (Figure 4b), the content of SS$_{bola}$C18:1 in a fiber environment at each temperature is: $X_F$= 100 % at 4°C, 60 % at 20°C, 10 % at 33°C and 0% at 50°C. The comparison between the liquid- and solid-state spectra shows that the peaks at 29.3 ppm (-(CH$_2$)$_n$-) and 129.5 ppm (9,10), both related to the aliphatic linker, and at 61.1 ppm (6',6''), 70.0 (4',4'',1,18) and 75.7 (2'',3',3'',5',5''), belonging to sophorose, resonate at the same chemical shifts. On the contrary, 2' (80-85 ppm range), located on the inner glucose group of sophorose, undergoes a significant downfield shift of ~1.5 ppm. Similarly, the position and relative intensity of sophorose anomeric carbons 1' and 1'' (100-106 ppm range) are also affected.

In terms of the relative intensity, solution NMR shows that both peaks must obviously have comparable intensity; however, in the solid-state, 1'' (104.2 ppm) becomes more intense when going from the micellar (50°C, $X_F$= 0%) to the fibrillar (4°C, $X_F$= 100%) solid. One should not disregard the fact that these experiments are performed under cross-polarization, meaning that they are semi-quantitative and comparison of the relative intensities should be done with care. However, the same nature of the chemical group (1'≡1''≡ CH), having the same position within the molecular skeleton (anomeric carbon in glucose) and use of the same cross-polarization time (1 ms) do allow a semi-quantitative interpretation. One can reasonably state that the intensity inversion is dominated by a different strength of the C-H dipolar coupling and relaxation time during spin lock ($t_{1\rho}$). For the same chemical group, differences in the strength of dipolar coupling and relaxation times are then generally explained by a different mobility and network of H-bonding interactions, whereas the more rigid C-H group and more extended H-bonding network enhance the intensity in the cross-polarization build-up profile.

Many conformational states are known for sophorose[31,32] but also for cyclic and linear carbohydrates in general.[33–35,70–73] Conformational changes of sugars are responsible for their physical and biochemical properties but also, in the case of hexonamide derivatives, crystal structures and phases.[36–38] Chemical shifts in the order of 1.5 ppm, as well as inversions in the signal intensity are directly connected to important conformational changes in disaccharides,[33,34] with strong implications in the phase behaviour.[36–38,72–74] The appreciable spectroscopic changes, found between the liquid and solid-state spectra, and across the solid-state spectra themselves, are observed mainly at the level of the 1'' and 2' carbons of sophorose. This indicates important conformational changes along the β(1,2) linkage, in agreement with the literature.[36–38,75]

A direct relationship between the $^{13}$C chemical shifts and a given conformation of the sugar headgroup of SS$_{bola}$C18:1 cannot be done, mainly due to the lack of a reference crystal structure, as otherwise discussed for alkyl hexonamides,[36–38] and known energy landscape for this compound. In the case of alkyl hexonamides, downfield shifts were attributed to *anti* and *trans* conformations, while *gauche* effects, in some cases up to the γ position in the sugar chain, induce upfield chemical shifts.[36–38,75] Energy landscapes are known for single disaccharides in glassy and solution state but not for most of their derivatives, including SS$_{bola}$C18:1. Complex two-dimensional maps correlate the energy landscape and $^{13}$C chemical shifts with the dihedral angles, φ and ψ, where φ=(O5''–C1''–O2'–C2') and ψ =(C2'–O2'–C1''–O5'') are defined along the β(1,2) glycosidic bond, with the convention of φ= 180°= *trans*, ψ= 180°= *trans* conformations (scheme in Figure 7).[32–34]

In light of the above, it could be tempting to identify a well-defined sophorose conformation on the basis of the $^{13}$C chemical shifts. However, in the absence of a reference crystal structure and molecular modelling, any tentative correlation would be highly speculative. Concerning the higher intensity of the 1'' peak at 104.2 ppm, two arguments could be invoked. The first one is based on cross polarization arguments. Conformational changes from micelles (50°C into fibers (4°C) result in the enhanced rigidity of 1'', with a more polarization transfer and higher intensity. The second one is based on the coexistence of two conformations, where one is in excess with respect to the other, as found for crystallized sophorose.[32]

*Sugar conformation drives the energy landscape of SS$_{bola}$C18:1 self-assembly*

Low-molecular weight SAFIN are long-studied for various classes of compounds, such as peptide amphiphiles, FMOC-derivatives or glycolipids.[9,25,62] The mechanisms behind the fibrillation process are complex, and hardly generalizable, because they result from a subtle equilibrium of intermolecular forces, such as Van der Waals attraction and electrostatic repulsion, but H-bonding and hydration effects may contribute.[54] Changes in the molecular conformation can play crucial roles in the equilibrium morphology, and in this regard the chemical composition of the fibrillating compound is crucial. Plus, other physical phenomena, like kinetically-driven supersaturation, can play an important role, as well.[5,29]

A more recent trend tried to reason out supramolecular fibrillation and gelation through the energy landscape approach. Specific molecular conformations are associated to a given supramolecular morphology like micelles, short or long fibers, nanotubes, each possibly thermodynamically stable or kinetically-trapped, according to the type of molecule and physicochemical conditions employed. The thermodynamic self-assembled form of SS$_{bola}$C18:1 are the fiber form below 14°C and the micellar aggregation above about 40°C, in which the sophorose headgroups are characterized by a different pair of dihedral angles, (φ,ψ) and (φ',ψ') respectively (Figure 7a). Nonetheless, the micellar morphology becomes the thermodynamic phase upon annealing above 40°C, but also kinetically-trapped over hours/days upon cooling, whichever the cooling rate and even below the fiber-to-micelle transition temperature, at about 14°C (Figure 7b).

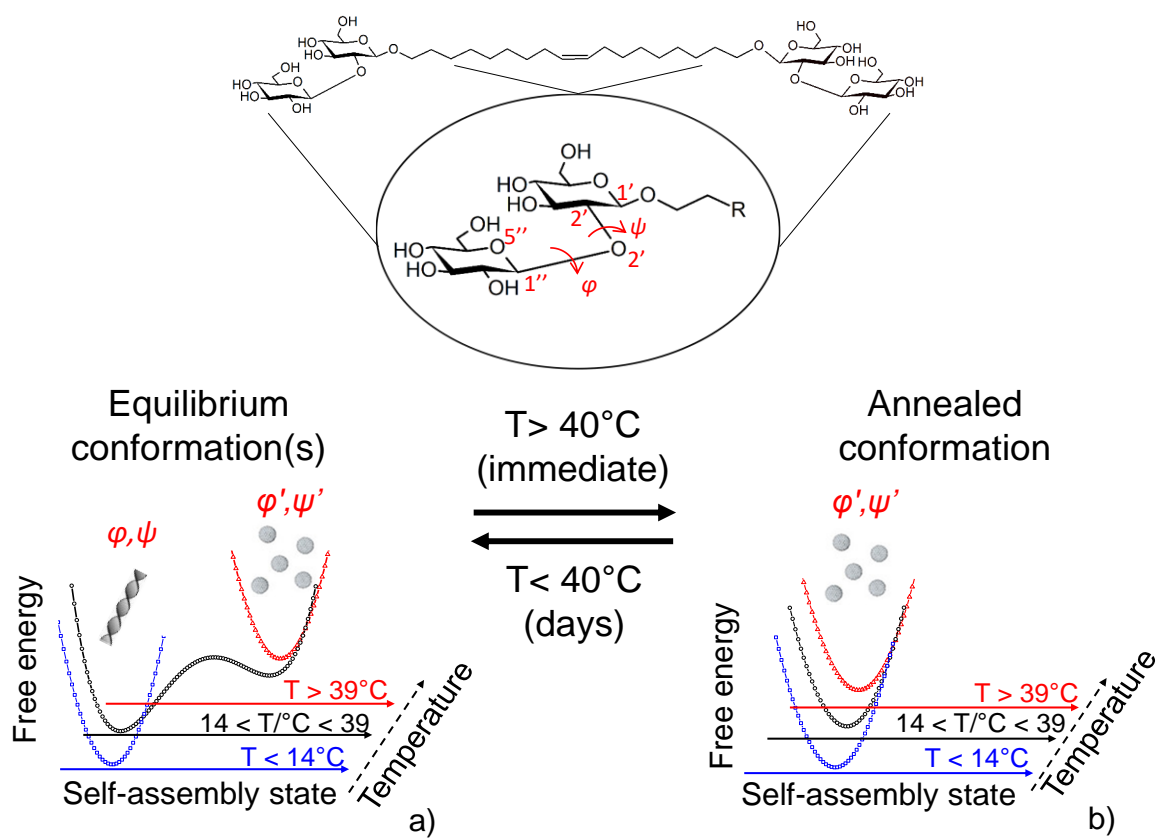

**Figure 7** – Schematic representation of the energy landscape of SS$_{bola}$C18:1 according to the conformation of the sophorose headgroup. Equilibrium conformations a) before and after b) annealing T> 40°C. φ and ψ are the O5''-C1''-O2'-C2' and C2'- O2'- C1''- O5'' dihedral angles along the β(1,2) glycosidic bond of sophorose.[32–34]

In analogy with peptide derivatives,[13] we propose that SS$_{bola}$C18:1 self-assembles into various stable structures, each characterized by a specific energy minimum. Each minimum is in turn associated to a specific conformation of sophorose, for which several energy minima are known as a function of φ and ψ,[31,32] in analogy to other disaccharides.[33,34,71] Since the values of φ and ψ corresponding to a minimized energy state in sophorose are not known for SS$_{bola}$C18:1, we propose in Figure 7, as reported for other systems,[13,14] a qualitative energy landscape for each self-assembled structure of SS$_{bola}$C18:1. Below the annealing temperature and at equilibrium, SS$_{bola}$C18:1 assembles into an orthogonal micelle-fiber network at room temperature (black, Figure 7a) and fully fiber network below 14°C (blue, Figure 7a), where sophorose is characterized by φ and ψ. Upon annealing, the dihedral angles change into φ' and ψ', thus favouring the micellar morphology. Considering the multiple minima in the energetic conformations of sophorose, we argue that the conformations of sophorose associated to (φ, ψ) and (φ', ψ') in SS$_{bola}$C18:1 are also energetically close, thus trapping sopohorose into the (φ', ψ') conformation for several hours, or days. The transition between the annealed and thermodynamic conformations seems to depend on the probability of conformational changes in the most external glucose moiety of sophorose.

Interestingly, both conformations seem to be stable even if the $SS_{bola}C18:1$ solution is dried or freeze-dried. We have observed that freeze-drying of the annealed, micellar, sol will equally provide a sol upon solubilization in water at a later stage. This could explain the micellar environment, previously characterized for this same compound upon dispersion in water.[76] On the contrary, freeze-drying of the fibrillar form will result in prompt fibrillation and hydrogel formation upon dispersion in water.

*General validity of the energy landscape approach for glycolipid biosurfactants*

Fibrillation has been reported for a number of glycolipid biosurfactants in the literature[21] and some specific molecular systems are able to form hydrogels[23,26,43] or organogels.[59] In light of the present results, one could suspect a general trend for this class of compounds. The symmetrical C16:0 bola sophoroside, $SS_{bola}C16:0$, undergoes a gel-to-sol transition under similar conditions.[43] If annealing effects were not specifically studied, authors did observe long kinetics of hydrogel formation upon cooling. Thus, $SS_{bola}C16:0$ could also be characterized by two energetically-close conformations of sophorose at room temperature. Other symmetrical and non-symmetrical bolaform sophorolipid and sophoroside analogues, under current study by our group, seem to show similar properties and could be worth studying in light of the findings observed for $SS_{bola}C18:1$.

The *trans* derivative of acidic sophorolipids, SL-C18:1$_{trans}$, was shown to fibrillate under acidic pH conditions,[77,78] while alkaline pH induces a fiber-to-micelle transition.[78] However, it was shown that fibrillation also occurs at basic pH with time (hours to days). In this case, annealing seems to occur by pH rather than by temperature.[78] On the contrary, on-the-shelf observations performed for acidic C18:0, C16:0 sophorolipids and cellobioselipids, all forming a fiber phase at acidic pH and room temperature,[26,46,57] show a readily reversible fiber-to-micelle-to-fiber transition over a heating and cooling cycle, thus indicating that annealing into a given sophorose or cellobiose conformation does not occur. These observations eventually demonstrate the so far unpredictable behaviour of this class of biobased glycolipids and the need to study them further to eventually achieve a properties control through molecular design for real-life applications.

As an overall conclusion, to the best of our knowledge, conformation-dependent hydrogelation is not reported for glycolipids and actually seldom reported for self-assembled fibrillary hydrogels in general,[11,12] despite the fact that conformation-dependent crystalline structures in glycolipids have been known for decades.[36–38] Considering the complexity of sugar conformation[33,35,71,79] and recent advances in applying the energy landscape approach to describe the self-assembly of complex amphiphiles,[13,14] it could be worth to dedicate future work to the structure-property relationship in glycosylated amphiphiles.

**Conclusions**

The symmetric bolaform sophoroside $SS_{bola}C18:1$ is an interesting compound obtained by microbial fermentation and belonging to the class of biosurfactants. Previously reported to form a stable micellar phase at room temperature, the self-assembly behaviour of this compound is much more complicated and it reveals a micelle-to-fiber, sol-to-gel, transition, dependent on the conformation of the double sophorose headgroup. Sophorose is a β(1,2)-D-glucose disaccharide known to have at least three low-energy conformations, each described by a pair of φ and ψ, the dihedral angles along the β(1,2) glycosidic bond. Cryo-TEM and SAXS show a stable low-temperature nanofibrillar and high-temperature micellar phases, which stabilize strong hydrogel and sol, respectively, as shown by oscillatory rheology and *in situ* rheo-SAXS.

The gel-to-sol-to-gel transition is readily reversible as long as temperature is below about 40°C. Above, the micellar phase becomes kinetically-trapped for hours, sometimes days, even if temperature is lowered (4°C) and independently on the cooling rate (1°C/min or 10°C/min). Chemical shift in $^{13}C$ NMR is very sensitive to little variations in the molecular conformation. Small (<1 ppm) to strong (up to 10-15 ppm) variation in the chemical shifts have been associated to the energetic landscape reflecting the conformation of disaccharides, both in solution and solid-state, and each having a specific pair of dihedral angles φ and ψ. A combination of solution and solid-state $^{13}C$ NMR shows that the 1'', 1' and 2' carbon stereocenters of sophorose are the only ones undergoing a chemical shift of at least 1 ppm, between the micellar and fibrillar phase, thus indicating two different pairs of φ and ψ for each phase. Although unknown for $SS_{bola}C18:1$, the energy landscape of sophorose and other disaccharides suggests that two or more conformations of comparable energy can coexist at the same time. We then attribute to this phenomenon the slow micelle-to-fiber, sol-to-gel, transition kinetics.

Coexistence of two or more carbohydrate conformations, each associated to a given crystal structure, is not uncommon in glycolipid amphiphiles but it is still poorly studied and never reported to natural glycolipids. Comparison between similar sophorolipids and sophorosides actually indicate that the phenomenon is neither general nor predictable, thus requiring more work on this family of compounds before employing them in real-life applications.

**Acknowledgements**

Authors kindly acknowledge the French ANR, Project N° SELFAMPHI - 19-CE43-0012-01. Cédric Lorthioir (Sorbonne Université, Paris, France) is kindly acknowledged for his assistance on the solid-state NMR experiments. ESRF synchrotron is acknowledged for financial support during the beamtimes associated to the proposal numbers N°MX-2311 and SC-4976.

**Supporting Information content**

Figure S 1 presents the strain and frequency dependent elastic and viscous moduli of SS$_{bola}$C18:1 gel at 2.5% and 1.25%. Figure S 2 shows the cryo-TEM images for SS$_{bola}$C18:1 solutions at 0.05 wt%. Figure S 3 presents the elastic and viscous moduli of SS$_{bola}$C18:1 gels under large strain. Figure S 4 shows the rheo-SAXS shear experiment

**References**

(1) Prince, E.; Kumacheva, E. Design and Applications of Man-Made Biomimetic Fibrillar Hydrogels. *Nat. Rev. Mater.* **2019**, *4*, 99–115.

(2) Ekiz, M. S.; Cinar, G.; Khalily, M. A.; Guler, M. O. Self-Assembled Peptide Nanostructures for Functional Materials. *Nanotechnology* **2016**, *27*, 1–37.

(3) Pape, A. C. H.; Dankers, P. Y. W. Supramolecular Hydrogels for Regenerative Medicine. *Adv. Polym. Sci.* **2015**, *268*, 253–279.

(4) Baglioni, P.; Carretti, E.; Chelazzi, D. Nanomaterials in Art Conservation. *Nat. Nanotechnol.* **2015**, *10*, 287–290.

(5) Yu, R.; Lin, N.; Yu, W.; Liu, X. Y. Crystal Networks in Supramolecular Gels: Formation Kinetics and Mesoscopic Engineering Principles. *CrystEngComm* **2015**, *17*, 7986–8010.

(6) Du, X.; Zhou, J.; Shi, J.; Xu, B. Supramolecular Hydrogelators and Hydrogels: From Soft Matter to Molecular Biomaterials. *Chem. Rev.* **2015**, *115*, 13165–13307.

(7) Hanabusa, K.; Suzuki, M. Development of Low-Molecular-Weight Gelators and Polymer-Based Gelators. *Polym. J.* **2014**, *46*, 776–782.

(8) Weiss, R. G. The Past, Present, and Future of Molecular Gels. What Is the Status of the Field, and Where Is It Going? *J. Am. Chem. Soc.* **2014**, *136*, 7519–7530.

(9) Draper, E. R.; Adams, D. J. Low-Molecular-Weight Gels: The State of the Art. *Chem* **2017**, *3*, 390–410.

(10) Menger, F. M.; Caran, K. L. Anatomy of a Gel. Amino Acid Derivatives That Rigidify Water at Submillimolar Concentrations. *J. Am. Chem. Soc.* **2000**, *122*, 11679–11691.

(11) Köhler, K.; Meister, A.; Förster, G.; Dobner, B.; Drescher, S.; Ziethe, F.; Richter, W.; Steiniger, F.; Drechsler, M.; Hause, G.; et al. Conformational and Thermal Behavior of a PH-Sensitive Bolaform Hydrogelator. *Soft Matter* **2006**, *2*, 77.

(12) Meister, A.; Bastrop, M.; Koschoreck, S.; Garamus, V. M.; Sinemus, T.; Hempel, G.; Drescher, S.; Dobner, B.; Richtering, W.; Huber, K.; et al. Structure-Property Relationship in Stimulus-Responsive Bolaamphiphile Hydrogels. *Langmuir* **2007**, *23*, 7715–7723.

(13) Tantakitti, F.; Boekhoven, J.; Wang, X.; Kazantsev, R. V.; Yu, T.; Li, J.; Zhuang, E.; Zandi, R.; Ortony, J. H.; Newcomb, C. J.; et al. Energy Landscapes and Functions of Supramolecular Systems. *Nat. Mater.* **2016**, *15*, 469–476.

(14) Fichman, G.; Guterman, T.; Damron, J.; Adler-Abramovich, L.; Schmidt, J.; Kesselman, E.; Shimon, L. J. W.; Ramamoorthy, A.; Talmon, Y.; Gazit, E. Supramolecular Chemistry: Spontaneous Structural Transition and Crystal Formation in Minimal Supramolecular Polymer Model. *Sci. Adv.* **2016**, *2*, e1500827–e1500827.

(15) Farias, C. B. B.; Almeida, F. C. G.; Silva, I. A.; Souza, T. C.; Meira, H. M.; Soares da Silva, R. de C. F.; Luna, J. M.; Santos, V. A.; Converti, A.; Banat, I. M.; et al. Production of Green Surfactants: Market Prospects. *Electron. J. Biotechnol.* **2021**, *51*, 28–39.

(16) Markande, A. R.; Patel, D.; Varjani, S. A Review on Biosurfactants: Properties, Applications and Current Developments. *Bioresour. Technol.* **2021**, *330*, 124963.

(17) Moutinho, L. F.; Moura, F. R.; Silvestre, R. C.; Romão-Dumaresq, A. S. Microbial Biosurfactants: A Broad Analysis of Properties, Applications, Biosynthesis, and Techno-Economical Assessment of Rhamnolipid Production. *Biotechnol. Prog.* **2021**, *37*, 1–14.

(18) Nikolova, C.; Gutierrez, T. Biosurfactants and Their Applications in the Oil and Gas Industry: Current State of Knowledge and Future Perspectives. *Front. Bioeng. Biotechnol.* **2021**, *9*, 626639.

(19) Shu, Q.; Lou, H.; Wei, T.; Liu, X.; Chen, Q. Contributions of Glycolipid Biosurfactants and Glycolipid-Modified Materials to Antimicrobial Strategy: A Review. *Pharmaceutics* **2021**, *13*, 227.

(20) Johnson, P.; Trybala, A.; Starov, V.; Pinfield, V. J. Effect of Synthetic Surfactants on the Environment and the Potential for Substitution by Biosurfactants. *Adv. Colloid Interface Sci.* **2021**, *288*.

(21) Baccile, N.; Seyrig, C.; Poirier, A.; Castro, S. A.; Roelants, S. L. K. W.; Abel, S. Self-

Assembly, Interfacial Properties, Interactions with Macromolecules and Molecular Modelling and Simulation of Microbial Bio-Based Amphiphiles (Biosurfactants). A Tutorial Review. *Green Chem.* **2021**, *23*, 3842–3944.

(22) Ben Messaoud, G.; Griel, P. Le; Prévost, S.; Merino, D. H.; Soetaert, W.; Roelants, S. L. K. W.; Stevens, C. V.; Baccile, N. Single-Molecule Lamellar Hydrogels from Bolaform Microbial Glucolipids. *Soft Matter* **2020**, *16*, 2528–2539.

(23) Ben Messaoud, G.; Le Griel, P.; Hermida-Merino, D.; Roelants, S. L. K. W.; Soetaert, W.; Stevens, C. V.; Baccile, N. PH-Controlled Self-Assembled Fibrillar Network (SAFiN) Hydrogels: Evidence of a Kinetic Control of the Mechanical Properties. *Chem. Mater.* **2019**, *31*, 4817–4830.

(24) Bettenhausen, C. BASF Invests in Biosurfactants. *C&EN* **2021**, *99*, 12.

(25) Johnson, E. K.; Adams, D. J.; Cameron, P. J. Peptide Based Low Molecular Weight Gelators. *J. Mater. Chem.* **2011**, *21*, 2024–2027.

(26) Baccile, N.; Messaoud, G. Ben; Griel, P. Le; Cowieson, N.; Perez, J.; Geys, R.; Graeve, M. De; Roelants, S. L. K. W.; Soetaert, W. Palmitic Acid Sophorolipid Biosurfactant: From Self-Assembled Fibrillar Network (SAFiN) To Hydrogels with Fast Recovery. *Philos. Trans. A* **2021**, *379*, 20200343.

(27) Colquhoun, C.; Draper, E. R.; Schweins, R.; Marcello, M.; Vadukul, D.; Serpell, L. C.; Adams, D. J. Controlling the Network Type in Self-Assembled Dipeptide Hydrogels. *Soft Matter* **2017**, *13*, 1914–1919.

(28) Johnson, E. K.; Adams, D. J.; Cameron, P. J. Directed Self-Assembly of Dipeptides to Form Ultrathin Hydrogel Membranes. *J. Am. Chem. Soc.* **2010**, *132*, 5130–5136.

(29) Li, J. L.; Liu, X. Y. Architecture of Supramolecular Soft Functional Materials: From Understanding to Micro-/Nanoscale Engineering. *Adv. Funct. Mater.* **2010**, *20*, 3196–3216.

(30) Van Renterghem, L.; Roelants, S. L. K. W.; Baccile, N.; Uyttersprot, K.; Taelman, M. C.; Everaert, B.; Mincke, S.; Ledegen, S.; Debrouwer, S.; Scholtens, K.; et al. From Lab to Market: An Integrated Bioprocess Design Approach for New-to-Nature Biosurfactants Produced by Starmerella Bombicola. *Biotechnol. Bioeng.* **2018**, *115*, 1195–1206.

(31) Dowd, M. K.; French, A. D.; Reilly, P. J. Conformational Analysis of the Anomeric Forms of Sophorose, Laminarabiose, and Cellobiose Using MM3. *Carbohydr. Res.* **1992**, *233*, 15–34.

(32) Andre, I.; Mazeau, K.; Taravel, F. R.; Tvaroska, I. Nmr and Molecular Modeling of Sophorose and Sophorotriose in Solution. *New J. Chem.* **1995**, *19*, 331–339.

(33) Lefort, R.; Bordat, P.; Cesaro, A.; Descamps, M. Exploring Conformational Energy Landscape of Glassy Disaccharides by Cross Polarization Magic Angle Spinning C13 NMR and Numerical Simulations. I. Methodological Aspects. *J. Chem. Phys.* **2007**, *126*.

(34) Lefort, R.; Bordat, P. Exploring Conformational Energy Landscape of Glassy Disaccharides by CPMAS 13C NMR and DFT/GIAO Simulations. II. Enhanced Molecular Flexibility in Amorphous. *Arxiv Prepr. cond-mat/ ...* **2006**, 0606408.

(35) Cheetham, N. W. H.; Dasgupta, P.; Ball, G. E. NMR and Modelling Studies of Disaccharide Conformation. *Carbohydr. Res.* **2003**, *338*, 955–962.

(36) Svenson, S.; Koening, J.; Fuhrhop, J. H. Crystalline Order in Probably Hollow Micellar Fibers of N-Octyl-D-Gluconamide. *J. Phys. Chem.* **1994**, *98*, 1022–1028.

(37) Svenson, S.; Kirste, B.; Fuhrhop, J. H. A CPMAS 13C NMR Study of Molecular Conformations and Disorder of N-Octylhexonamides in Microcrystals and Supramolecular Assemblies. *J. Am. Chem. Soc.* **1994**, *116*, 11969–11975.

(38) Svenson, S.; Schaefer, A.; Fuhrhop, J. H. Conformational Effects of 1.3-Syn-Diaxial Repulsion and 1.2-Gauche Attraction Between Hydroxy Groups n Monomolecula r N-Octyl-D-Hexonamide Solutions A 13C and 'H NMR Spectroscopic Study Sonke. *J. Chem. Soc. Perkin Trans. 2* **1994**, *2*, 1023–1028.

(39) Schindelin, J.; Arganda-Carreras, I.; Frise, E.; Kaynig, V.; Longair, M.; Pietzsch, T.; Preibisch, S.; Rueden, C.; Saalfeld, S.; Schmid, B.; et al. Fiji: An Open-Source Platform for Biological-Image Analysis. *Nat. Methods* **2012**, *9*, 676–682.

(40) Narayanan, T.; Sztucki, M.; Zinn, T.; Kieffer, J.; Homs-Puron, A.; Gorini, J.; Van Vaerenbergh, P.; Boesecke, P. Performance of the Time-Resolved Ultra-Small-Angle X-Ray Scattering Beamline with the Extremely Brilliant Source. *J. Appl. Crystallogr.* **2022**, *55*, 98–111.

(41) Ashiotis, G.; Deschildre, A.; Nawaz, Z.; Wright, J. P.; Karkoulis, D.; Picca, F. E.; Kieffer, J. The Fast Azimuthal Integration Python Library: PyFAI. *J. Appl. Crystallogr.* **2015**, *48*, 510–519.

(42) de Gennes, P.-G. P.-G. Dynamics of Entangled Polymer Solutions. I. The Rouse Model. *Macromolecules* **1976**, *9*, 587–593.

(43) Baccile, N.; Renterghem, L. Van; Griel, P. Le; Ducouret, G.; Brennich, M.; Cristiglio, V.; Roelants, S. L. K. W.; Soetaert, W. Bio-Based Glyco-Bolaamphiphile Forms a Temperature-Responsive Hydrogel with Tunable Elastic Properties. *Soft Matter* **2018**, *14*, 7859–7872.

(44) Teixeira, J. Small-Angle Scattering by Fractal Systems. *J. Appl. Crystallogr.* **1988**, *21*,

781–785.

(45) Stradner, A.; Glatter, O.; Schurtenberger, P. Hexanol-Induced Sphere-to-Flexible Cylinder Transition in Aqueous Alkyl Polyglucoside Solutions. *Langmuir* **2000**, *16*, 5354–5364.

(46) Cuvier, A.-S. S.; Berton, J.; Stevens, C. V; Fadda, G. C.; Babonneau, F.; Van Bogaert, I. N. A. a; Soetaert, W.; Pehau-Arnaudet, G.; Baccile, N. PH-Triggered Formation of Nanoribbons from Yeast-Derived Glycolipid Biosurfactants. *Soft Matter* **2014**, *10*, 3950–3959.

(47) Masuda, M.; Shimizu, T. Lipid Nanotubes and Microtubes: Experimental Evidence for Unsymmetrical Monolayer Membrane Formation from Unsymmetrical Bolaamphiphiles. *Langmuir* **2004**, *20*, 5969–5977.

(48) Qiao, Y.; Lin, Y.; Wang, Y.; Yang, Z.; Liu, J.; Zhou, J.; Yan, Y.; Huang, J. Metal-Driven Hierarchical Self-Assembled One-Dimensional Nanohelices. *Nano Lett.* **2009**, *9*, 4500–4504.

(49) Abel, S.; Dupradeau, F.-Y.; Raman, E. P.; MacKerell, A. D.; Marchi, M. Molecular Simulations of Dodecyl-Beta-Maltoside Micelles in Water: Influence of the Headgroup Conformation and Force Field Parameters RID E-1792-2011. *J. Phys. Chem. B* **2011**, *115*, 487–499.

(50) Tanford, C. *The Hydrophobic Effect: Formation of Micelles and Biological Membranes*; John Wiley & Sons Inc: New York, 1973.

(51) Nagarajan, R. Self-Assembly of Bola Amphiphiles. *Chem. Eng. Commun.* **1987**, *55*, 251–273.

(52) Manet, S.; Cuvier, A. S.; Valotteau, C.; Fadda, G. C.; Perez, J.; Karakas, E.; Abel, S.; Baccile, N. Structure of Bolaamphiphile Sophorolipid Micelles Characterized with SAXS, SANS, and MD Simulations. *J. Phys. Chem. B* **2015**, *119*, 13113–13133.

(53) Zhou, S.; Xu, C.; Wang, J.; Gao, W.; Akhverdiyeva, R.; Shah, V.; Gross, R. Supramolecular Assemblies of a Naturally Derived Sophorolipid. *Langmuir* **2004**, *20*, 7926–7932.

(54) Barclay, T. G.; Constantopoulos, K.; Matisons, J. Nanotubes Self-Assembled from Amphiphilic Molecules via Helical Intermediates. *Chem. Rev.* **2014**, *114*, 10217–10291.

(55) Cui, H.; Muraoka, T.; Cheetham, A. G.; Stupp, S. I. Self-Assembly of Giant Peptide Nanobelts. *Nano Lett.* **2009**, *9*, 945–951.

(56) Cuvier, A. S.; Babonneau, F.; Berton, J.; Stevens, C. V.; Fadda, G. C.; Genois, I.; Le Griel, P.; Péhau-Arnaudet, G.; Baccile, N. Synthesis of Uniform, Monodisperse,

Sophorolipid Twisted Ribbons. *Chem. - An Asian J.* **2015**, *10*, 2419–2426.

(57) Baccile, N.; Selmane, M.; Le Griel, P.; Prévost, S.; Perez, J.; Stevens, C. V.; Delbeke, E.; Zibek, S.; Guenther, M.; Soetaert, W.; et al. PH-Driven Self-Assembly of Acidic Microbial Glycolipids. *Langmuir* **2016**, *32*, 6343–6359.

(58) Imura, T.; Yamamoto, S.; Yamashita, C.; Taira, T.; Minamikawa, H.; Morita, T.; Kitamoto, D. Aqueous Gel Formation from Sodium Salts of Cellobiose Lipids. *J. Oleo Sci.* **2014**, *63*, 1005–1010.

(59) Imura, T.; Kawamura, D.; Ishibashi, Y.; Morita, T.; Sato, S.; Fukuoka, T.; Kikkawa, Y.; Kitamoto, D. Low Molecular Weight Gelators Based on Biosurfactants, Cellobiose Lipids by Cryptococcus Humicola. *J. Oleo Sci.* **2012**, *61*, 659–664.

(60) Clemente, J.; Romero, P.; Serrano, J. L.; Fitremann, J.; Oriol, L. Supramolecular Hydrogels Based on Glycoamphiphiles: E Ff Ect of the Disaccharide Polar Head. **2012**, No. Lc.

(61) Jung, J. H.; John, G.; Masuda, M.; Yoshida, K.; Shinkai, S.; Shimizu, T. Self-Assembly of a Sugar-Based Gelator in Water: Its Remarkable Diversity in Gelation Ability and Aggregate Structure. *Langmuir* **2001**, *17*, 7229–7232.

(62) Terech, P.; Weiss, R. G. Low Molecular Mass Gelators of Organic Liquids and the Properties of Their Gels. *Chem. Rev.* **1997**, *97*, 3133–3159.

(63) Chen, L.; Raeburn, J.; Sutton, S.; Spiller, D. G.; Williams, J.; Sharp, J. S.; Griffiths, P. C.; Heenan, R. K.; King, S. M.; Paul, A.; et al. Tuneable Mechanical Properties in Low Molecular Weight Gels. *Soft Matter* **2011**, *7*, 9721.

(64) Graf, G.; Drescher, S.; Meister, A.; Garamus, V. M.; Dobner, B.; Blume, A. Tuning the Aggregation Behaviour of Single-Chain Bolaamphiphiles in Aqueous Suspension by Changes in Headgroup Asymmetry. *Soft Matter* **2013**, *9*, 9562–9571.

(65) Adams, D. J.; Butler, M. F.; Frith, W. J.; Kirkland, M.; Mullen, L.; Sanderson, P. A New Method for Maintaining Homogeneity during Liquid–Hydrogel Transitions Using Low Molecular Weight Hydrogelators. *Soft Matter* **2009**, *5*, 1856.

(66) Fuhrhop, J. H.; Svenson, S.; Boettcher, C.; Rössler, E.; Vieth, H. M. Long-Lived Micellar N-Alkylaldonamide Fiber Gels. Solid-State NMR and Electron Microscopic Studies. *J. Am. Chem. Soc.* **1990**, *112*, 4307–4312.

(67) Feio, G.; Cohen-Addad, J. P. NMR Approach to the Kinetics of Polymer Crystallization. 1. Cis-1,4-Polybutadiene. *J. Polym. Sci. Part B Polym. Phys.* **1988**, *26*, 389–412.

(68) Esch, J. H. Van. We Can Design Molecular Gelators , But Do We Understand Them ? †. **2009**, *25*, 8392–8394.

(69) Lescanne, M.; Colin, A.; Mondain-Monval, O.; Fages, F.; Pozzo, J. L. Structural Aspects of the Gelation Process Observed with Low Molecular Mass Organogelators. *Langmuir* **2003**, *19*, 2013–2020.

(70) Bock, K.; Pedersen, C. Carbon-13 Nuclear Magnetic Resonance Secstroscopy of Monosaccharides. *Adv. Carbohydr. Chem. Biochem.* **1983**, *41*, 27–66.

(71) Biarnés, X.; Ardèvol, A.; Planas, A.; Rovira, C.; Laio, A.; Parrinello, M. The Conformational Free Energy Landscape of β-D-Glucopyranose. Implications for Substrate Preactivation in β-Glucoside Hydrolases. *J. Am. Chem. Soc.* **2007**, *129*, 10686–10693.

(72) Kirschner, K. N.; Woods, R. J. Solvent Interactions Determine Carbohydrate Conformation. *Proc. Natl. Acad. Sci. U. S. A.* **2001**, *98*, 10541–10545.

(73) Tvaroška, I.; Taravel, F. R.; Utille, J. P.; Carver, J. P. Quantum Mechanical and NMR Spectroscopy Studies on the Conformations of the Hydroxymethyl and Methoxymethyl Groups in Aldohexosides. *Carbohydr. Res.* **2002**, *337*, 353–367.

(74) Kroon-Batenburg, L. M. J.; Kroon, J. Solvent Effect on the Conformation of the Hydroxymethyl Group Established by Molecular Dynamics Simulations of Methyl-β-D-glucoside in Water. *Biopolymers* **1990**, *29*, 1243–1248.

(75) Sack, I.; Macholl, S.; Fuhrhop, J. H.; Buntkowsky, G. Conformational Studies of Polymorphic N-Octyl-D-Gluconamide with 15N (Labeled) 13C (Natural Abundance) REDOR Spectroscopy. *Phys. Chem. Chem. Phys.* **2000**, *2*, 1781–1788.

(76) Roelants, S. L. K. W.; Renterghem, L. Van; Maes, K.; Everaert, B.; Redant, E.; Vanlerberghe, B.; Demaeseneire, S.; Soetaert, W. Taking Biosurfactants from the Lab to the Market: Hurdles and How to Take Them by Applying an Integrated Process Design Approach. In *Microbial Biosurfactants and their Environmental and Industrial Applications*; Banat, I. M., Thavasi, R., Eds.; CRC Press, 2018.

(77) Dhasaiyan, P.; Banerjee, A.; Visaveliya, N.; Prasad, B. L. V. Influence of the Sophorolipid Molecular Geometry on Their Self-Assembled Structures. *Chem. Asian J.* **2013**, *8*, 369–372.

(78) Dhasaiyan, P.; Prévost, S.; Baccile, N.; Prasad, B. L. V. PH- and Time-Resolved in-Situ SAXS Study of Self-Assembled Twisted Ribbons Formed by Elaidic Acid Sophorolipids. *Langmuir* **2018**, *34*, 2121–2131.

(79) Alonso, J. L.; Lozoya, M. A.; Peña, I.; López, J. C.; Cabezas, C.; Mata, S.; Blanco, S. The Conformational Behaviour of Free D-Glucose - At Last. *Chem. Sci.* **2014**, *5*, 515–522.